\documentclass[sigconf, authorversion]{acmart} 

\setcopyright{none}              
\renewcommand\footnotetextcopyrightpermission[1]{} 

\usepackage{caption}
\usepackage{svg}
\usepackage{graphicx}

\renewenvironment{quote}
  {\list{}{\leftmargin=2em \rightmargin=2em}\item\relax}
  {\endlist}
  
\setcopyright{acmlicensed}
\copyrightyear{2025}
\acmYear{2025}
\acmDOI{XX.XX}

\acmConference[Future Conference]{Future Conference}{2025}{USA}

\acmISBN{978-1-4503-XXXX-X/18/06}

\begin{document}

\title[Exploring Student Behaviors \& Motivations using AI TAs with Optional Guardrails]{Exploring Student Behaviors and Motivations when using AI Teaching Assistants with Optional Guardrails}



\author{Amanpreet Kapoor}
\orcid{0000-0003-1340-8315}
\affiliation{%
  \institution{University of Florida}
  \city{Gainesville}
  \state{FL}
  \country{United States}
}
\email{kapooramanpreet@ufl.edu}

\author{Marc Diaz}
\orcid{0009-0009-3400-8362}
\affiliation{%
  \institution{University of Florida}
  \city{Gainesville}
  \state{FL}
  \country{United States}
  }
\email{marcgabe15@gmail.com}

\author{Stephen	MacNeil}
\orcid{0000-0003-2781-6619}
\affiliation{
  \institution{Temple University}
  \city{Philadelphia}
  \state{PA}
  \country{United States}}
\email{stephen.macneil@temple.edu}

\author{Leo Porter}
\orcid{0000-0003-1435-8401}
\affiliation{
  \institution{University of California, San Diego}
  \city{San Diego}
  \state{CA}
  \country{United States}
}
\email{leporter@ucsd.edu}

\author{Paul Denny}
\orcid{0000-0002-5150-9806}
\affiliation{
  \institution{University of Auckland}
  \city{Auckland}
  \country{New Zealand}
}
\email{paul@cs.auckland.ac.nz}

\begin{abstract}

AI-powered chatbots and digital teaching assistants (AI TAs) are gaining popularity in programming education, offering students timely and personalized feedback. Despite their potential benefits, concerns about student over-reliance and academic misconduct have prompted the introduction of ``guardrails'' into AI TAs---features that provide scaffolded support rather than direct solutions. However, overly restrictive guardrails may lead students to bypass these tools and use unconstrained AI models, where interactions are not observable, thus limiting our understanding of students' help-seeking behaviors. To investigate this, we designed and deployed a novel AI TA tool with optional guardrails in one lab of a large introductory programming course. As students completed three code writing and debugging tasks, they had the option to receive guardrailed help or use a ``See Solution'' feature which disabled the guardrails and generated a verbatim response from the underlying model. We investigate students' motivations and use of this feature and examine the association between usage and their course performance. We found that 50\% of the 885 students used the ``See Solution'' feature for at least one problem and 14\% used it for all three problems. Additionally, low-performing students were more likely to use this feature and use it close to the deadline as they started assignments later. The predominant factors that motivated students to disable the guardrails were assistance in solving problems, time pressure, lack of self-regulation, and curiosity. Our work provides insights into students' solution-seeking motivations and behaviors, which has implications for the design of AI TAs that balance pedagogical goals with student preferences.

\end{abstract}

\ccsdesc[500]{Social and professional topics~Computing education}

\keywords{AI tutor, Feedback, Digital TAs, Automated Tutors, Programming}

\maketitle

\section{Introduction}

Providing students with timely and high-quality help is essential in programming courses.  However, scaling the support that can be provided by human teaching assistants (TAs) is difficult in large classes where the needs of students seeking help can vary considerably.  Thus, there has been increasing interest in the use of AI TAs for providing equitable support to learners in introductory courses. Recent work has shown that such tools can generate responses that are both correct and helpful \cite{denny_ai_ta_desireable}, provide students with the sense of having a personal tutor \cite{liu2024cs50}, and produce timely, tailored feedback \cite{liffiton_codehelp}.  However, the growing adoption of AI TAs has also raised concerns such as inaccuracies in the AI-generated content \cite{denny_computing_education_cacm}, student over-reliance on such tools \cite{denny_computing_education_cacm, vadaparty2024cs1} which can potentially compound metacognitive difficulties \cite{prather2024widening}, and the risk that students may generate and copy solutions instead of learning concepts \cite{lau_ban_it}.



To address concerns about over-reliance and misuse, researchers have explored incorporating `guardrails' into AI TAs to provide scaffolded support that encourages learning rather than offering direct answers \cite{liffiton_codehelp, denny_ai_ta_desireable}. While students often appreciate such safeguards, they may also seek direct solutions when under time pressure, highlighting the complex dynamics of help-seeking behaviors \cite{krause2022exploration, lim2023student}. If the guardrails are perceived as overly restrictive, students may turn to unconstrained tools like ChatGPT, which lack pedagogical safeguards and instructor oversight and can undermine learning goals. Understanding how students interact with and value these systems is thus critical to designing digital TAs that balance educational objectives with student preferences.

To explore student-interactions with guardrailed AI TAs further, we conducted an observational study in a large introductory programming course.  We built and deployed a novel AI TA with \emph{optional} guardrails in our Edugator tool \cite{edugator_2024}, and asked students to complete three programming tasks---focused on code writing and debugging---while interacting with the AI TA. Students could choose whether to use the guardrailed version, which would not reveal code solutions, or disable the restrictions via a ``See Solution'' feature to generate a verbatim response from the underlying model which may include a code solution.  Once they had completed the tasks, the students provided feedback about the tool and described their motivations for using (or not using) the ``See Solution'' feature.  We investigate interaction patterns and performance based on the use of this feature, guided by the following research questions:

\begin{enumerate}
     \item[RQ.1a] To what extent is the ``See Solution'' feature used, and how does this usage relate to student performance in the course?
     \item[RQ.1b] How does the timing of student engagement with the lab tasks relate to the use of the ``See Solution'' feature?
    \item[RQ.2 ] What factors motivate students to use or refrain from using the ``See Solution'' feature when engaging with an AI TA?    
\end{enumerate}





\section{Related Work}

\subsection{Help-Seeking in Computing Courses}

Effective help-seeking is an important metacognitive skill \cite{aleven2006metacog}, especially for novice programmers who can obtain help in many ways, but may engage in unproductive behaviors like avoiding or over-relying on help \cite{marwan2020unproductive, skripchuk2023analysis}. 
Human TAs have traditionally provided front-line support when students are stuck, although prior research has revealed mismatches between student expectations and educational goals during help-seeking interactions with TAs.  In a recent study by Krause-Levy et al., researchers observed that TAs frequently provide solutions during help sessions rather than guide students through the problem-solving process \cite{krause2022exploration}. Follow on interview studies with students and TAs found that student incentives \cite{lim2023student} and TA incentives \cite{molina2025undergraduate} often support seeking or receiving a solution.

In addition, it can be challenging for tutors to spend the necessary time with each student who seeks help when handling multiple requests at busy times \cite{ riese2021challenges,liao2019behaviors, markel2021inside}. 
This can be exacerbated by student procrastination, which is a well-known barrier to success in programming courses \cite{edwards2009comparing}.  
Students who delay starting assignments not only risk poorer academic outcomes but also leave insufficient time to seek meaningful help \cite{zhang2022procrastination}.
The increased demand for help close to deadlines can place a significant strain on TAs, who are already limited in their availability outside typical working hours.  Moreover, not all students feel equally confident to seek help in-person \cite{smith2017my}.  These challenges have driven an interest in more consistent, scalable and private support options, such as digital AI TAs. 

\subsection{Emergence of AI-Powered Assistants}

The rapid improvement in capabilities of large language models (LLMs) has enabled the creation of AI-powered digital teaching assistants (AI TAs).  Initial work in this space evaluated the quality of responses produced by open-source and proprietary LLMs to historical repositories of student questions, finding generally good results \cite{ hellas2023exploring, koutcheme2024open, hicke2023aita}.
More recently, tools have been developed with integrated LLM support and deployed to answer student queries in real-time \cite{liu2024cs50}.
However, ensuring that responses are pedagogically sound remains a challenge.  Qiao et al. address this by providing instructor oversight of the responses generated by the LLM, but this limits scalability \cite{qiao2024oversightactionexperiencesinstructormoderated}.  More commonly, AI-powered tutoring tools are being designed to provide appropriate help automatically so that students can call on them when needed \cite{ma2025integrating}.

A common approach to prevent misuse when designing AI TAs is the use of `guardrails'.  Examples are implemented in tools like CodeHelp \cite{liffiton_codehelp} and CodeAid \cite{kazemitabaar_codeaid} which offer scaffolded support instead of producing direct code solutions.  Interestingly, students often say they prefer such scaffolded tools \cite{denny_ai_ta_desireable}.
\emph{However, if the guardrails in an AI TA are perceived as overly restrictive, students may turn to unconstrained LLMs like ChatGPT, which lack pedagogical safeguards}.  Such interactions are not monitored, and thus valuable insights into student learning behaviors are lost.  Understanding how students interact with and value AI TAs is essential for addressing these challenges and ensuring their integration meets learning objectives.

\begin{figure*}[ht!]
    \centering
    \captionsetup{justification=centering}
    \includegraphics[width=1\linewidth]{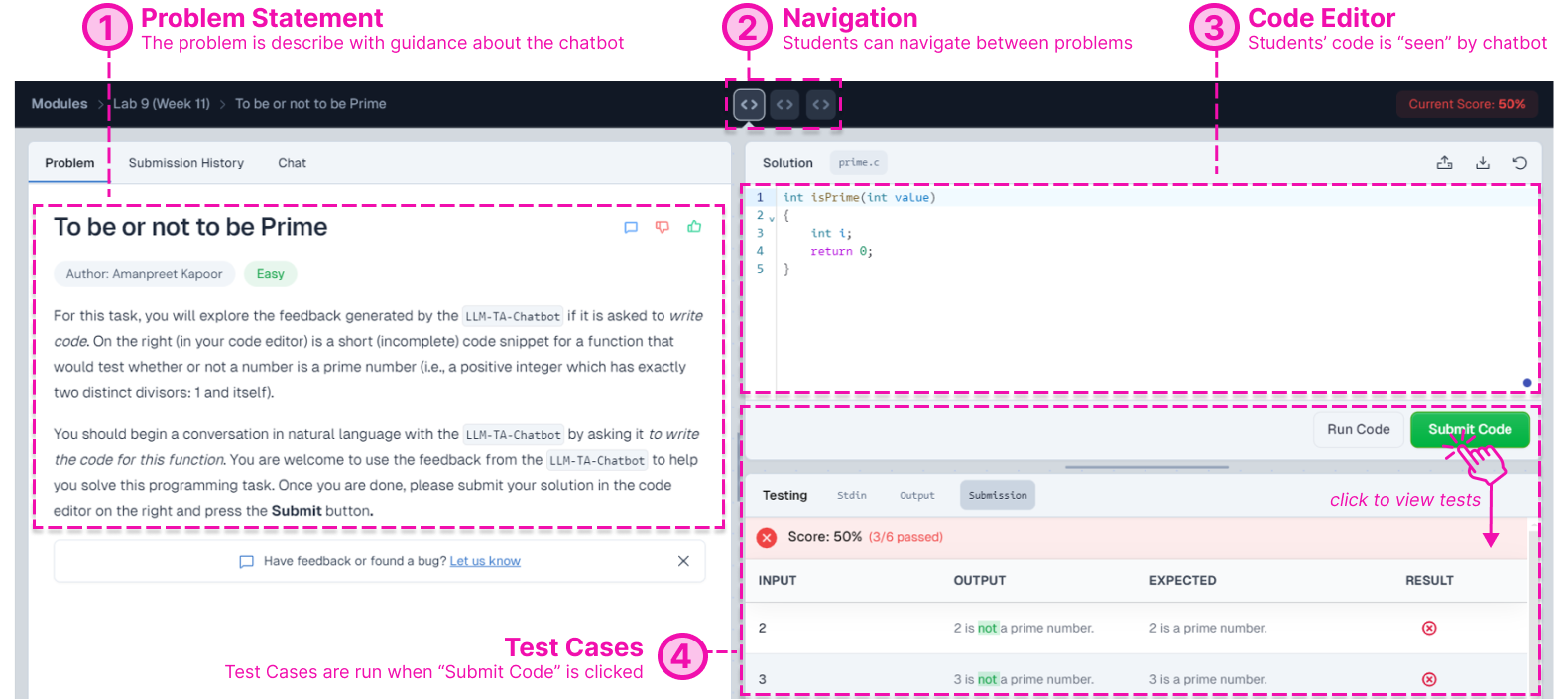}
    \caption{Interface for Edugator Tool and Lab Activity 1: \textit{isPrime} Code Writing Task}
    \label{fig:problem}
    \Description{This image describes the Anonymous Tool interface consisting of problem description, problem navigation, code editor and immediate feedback when students submit code}
\end{figure*}

\subsection{Over-Reliance on GenAI}

One of the most commonly cited concerns regarding GenAI use in CS education is that of student over-reliance \cite{denny_computing_education_cacm,prather2024beyond,vadaparty2024cs1, franklin2025generative}. Tools like ChatGPT and GitHub Copilot can generate detailed solutions to many programming-related tasks, which may undermine the development of problem-solving and computational thinking skills \cite{zastudil2023generative}. Lau and Guo warn that such reliance can harm foundational learning, while Sheard et al. highlight the challenges GenAI poses to maintaining academic integrity \cite{lau_ban_it, sheard2024instructor}.  In addition, students themselves recognize the risks of over-reliance. Hou et al. found that while students value the convenience of GenAI, they acknowledge that improper use can harm independent learning \cite{hou2024effects}. Despite this, GenAI tools are frequently used. Nearly a quarter of surveyed students reported using tools like ChatGPT daily or hourly, often for debugging and code generation \cite{hou2024effects}. 


The risks associated with over-reliance grow as GenAI tools advance. Recent studies have demonstrated that tools like GPT-4 can solve a wide range of programming problems, including visual challenges like Parsons Problems and computer graphics tasks \cite{hou2024more, gutierrez2024seeing, feng2024more}.
While this capability can accelerate problem-solving for advanced students, Prather et al. caution that struggling students may adopt ``shepherding'' behaviors, relying heavily on AI without meaningful engagement \cite{prather2024widening}. This reliance may create an illusion of competence, leaving essential skills underdeveloped.

The challenge lies in balancing the need for pedagogical safeguards with the flexibility to meet students' immediate help-seeking preferences. 
In this study, we examine student interaction with an AI TA that incorporates scaffolded support through guardrails.  
It also includes a feature that allows students, when needed, to remove these restrictions to access unfiltered LLM-generated responses.  This dual approach aims to balance pedagogical safeguards with student preferences, and provide insight into how such features influence help-seeking behaviors.

\section{Methods}

\subsection{Study Design and Context}
Our study was conducted in the context of a large introductory programming course offered at 
a large public research university in the Australasian region in Fall 2024. Data collected in this study was covered by an ethics approval for the analysis of naturally occurring coursework data.
A total of 1,034 students were enrolled in the course, which spanned a 12-week teaching term.  Students completed weekly lab exercises, each contributing 1\% toward their final grade.  Data for this study was collected in one of the weekly labs (described in Section \ref{lab}) where students were asked to complete one code writing task and two code debugging tasks (using the C language), and complete two quantitative and two qualitative questions. This lab was graded as part of the coursework.

The lab tasks were hosted on our novel tool (described in Section \ref{tool}) which offered students access to an AI chatbot with optional guardrails. All enrolled students were sent an invitation to participate (909 students accepted the invitation to the tool; 885 completed at least one of the three programming tasks). In this paper, we analyze data collected from the 885 participating students. 


\subsection{Tool} 
\label{tool}

\begin{figure}[ht!]
    \centering
    \captionsetup{justification=centering}
    \includegraphics[width=1.0\linewidth]{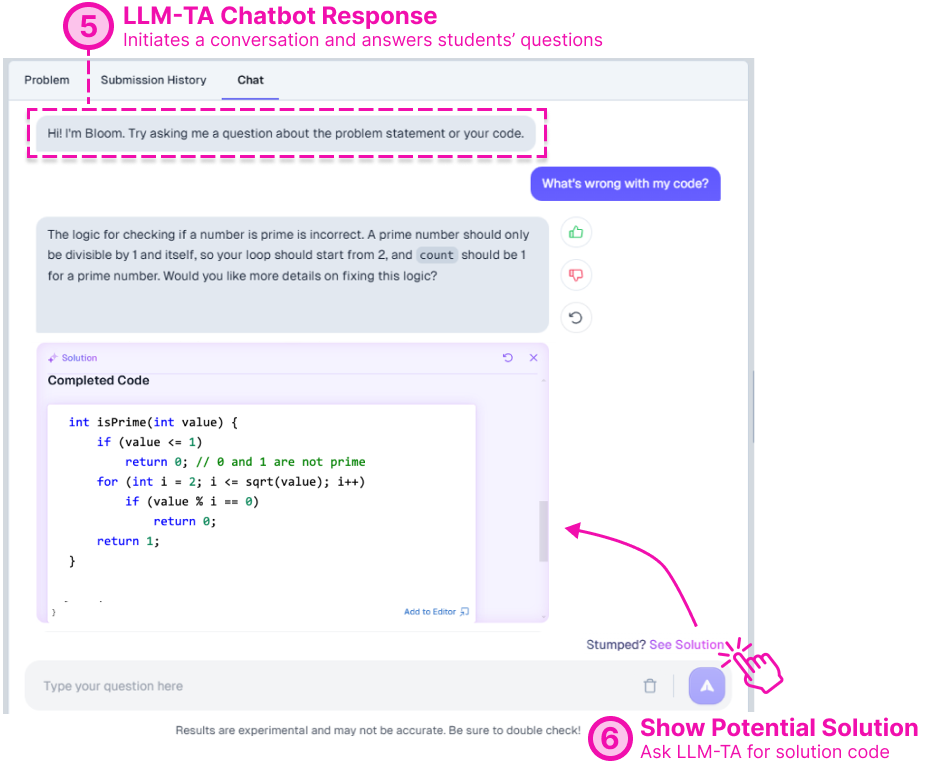}
    \caption{Chat interface to ask questions and solicit solution}
    \label{fig:solution}
    \Description{This image describes the Edugator Tool interface consisting of the AI TA chatbot and the ``See Solution'' button to retrieve the solution from the Chatbot}
\end{figure}

We developed the Edugator web-based tool \cite{edugator_2024} that allows instructors to write programming problems and offers students a clean interface to preview the problem description, code their solution in the browser, and receive immediate feedback on their solutions (see Figure \ref{fig:problem}). Additionally, the tool provides access to an LLM-powered AI TA (see Figure \ref{fig:solution}). This TA employs guardrails by way of a prompt that instructs the model to assist the student seeking help by explaining concepts or language syntax, 
similar to CodeHelp \cite{liffiton_codehelp}. The LLM is instructed not to provide any code snippets or a corrected version of a student's code solution. The prompt is fed the students' chat prompt, chat history, course programming language, and students' solution in the code editor. The prompt was not fed the problem description to ensure that students develop prompting skills by providing more context to the AI TA chatbot.

To evaluate solution-seeking behaviors for the purposes of this study, we added a new feature to the AI TA where students have the option to press the ``See Solution'' button after initiating the chat which allows them to deactivate the guardrails to see LLM generated-solution code along with the scaffolded response (see Figure \ref{fig:solution}). The button was labeled ``See Solution'' so that students were aware the feature can generate a solution, as they were unlikely to be familiar with the `guardrail' terminology used in literature. The prompt fed to retrieve the solution, asked the LLM to assume that it is \textit{``a coding expert tasked with providing solutions''}. Further, the model was instructed to provide the student with a detailed explanation of the problem, including the correct code solution. The problem statement was also sent to this prompt in addition to the aforementioned attributes that were used for the AI TA with guardrails. Our AI TA was backed by OpenAI's 4o model and the total cost of LLM usage during the study (9 days) was US\$53.00.  

\subsection{Lab Activities} \label{lab}

For all lab tasks, the course used the CodeRunner platform \cite{lobb2016coderunner}. For this study, which was run as part of ``Lab 9'' in the course, students were provided some instructions within CodeRunner as well as a link to the Edugator tool \cite{edugator_2024}. The instructions stated that in order to earn the 1 mark (out of the lab's 10 marks) for the task, they needed to solve the three tasks in the Edugator tool. These programming tasks corresponded to the course content covered in Week 9 (primarily targeting nested loops and two-dimensional arrays). All students saw these instructions to the tasks prior to clicking through to the Edugator tool. Edugator was demonstrated in class where attendance was optional, but all lectures were recorded and available online. So every student had access to a recording of this demonstration.  During this demonstration, it was made clear to students that they were welcome to make use of any of the features of  Edugator, including the ``See solution'' feature. Within the Edugator tool itself, for each problem description, students could see messages such as: ``\textit{You are welcome to use the feedback from the LLM-TA Chatbot to help you solve this programming task}''. 

As with all labs in the course, given the large size of the course, the labs were run over a period of one week. Students were free to complete the lab at any time during that week (online), and they could also attend an in-person lab (which is optional) that was scheduled throughout the week, if they wanted assistance from a human TA.

The open-ended and quantitative questions were configured as separate questions within CodeRunner and completing these questions allowed students to get 1 mark (out of the 10 marks). These questions were ordered to appear after the CodeRunner question that linked to the Edugator tool. It would be possible for students to skip ahead to the reflection question without completing the Edugator task. However, given the high number of students that both completed the Edugator task (n=885) and gave an open reflection response (n=839), we expect that most students did these in order (there is not much benefit or incentive for students to do them out of order). 
We now briefly describe the three activities.
 
\textbf{Activity 1: IsPrime (Code Writing)}. Activity 1 required students to finish the incomplete IsPrime() function provided (as shown in Figure \ref{fig:problem}). Students were prompted to \emph{``begin a conversation in natural language with the LLM TA Chatbot by asking it to write the code for this function''}.  Such a request would not result in a direct code solution, due to the presence of the guardrails.

\textbf{Activity 2: IsRepeated (Code Debugging)}. Activity 2 provided a buggy implementation of an IsRepeated() function, which should return true if a one-dimensional array contains any repeated values.  A nested loop was incorrectly defined by initialising the inner loop variable ($j$) to be equal to ($=i$) rather than larger than ($=i+1$) the value of the outer loop variable.  Students were asked to: \emph{``Copy the function (which has a bug) and provide it as input to the TA chatbot. You should construct a short natural language description explaining what the function intends to do and asking it to debug the code''}. Students were encouraged to follow this step, even if they could see the bug, so that they could view and critique the chatbot output.

\textbf{Activity 3: SurroundingSum (Code Debugging)}.  Activity 3 involved computing the sum of all values surrounding a specified location in a two-dimensional array.  The provided code contained a bug by failing to guard against out-of-bounds array accesses in the case that the location was on the border of the array.  Similar instructions were provided as for Activity 2.

\subsection{Data Collection and Analysis}
\subsubsection{Quantitative Data}
We analyze log data of student interactions with our tool, and summarize the activity in two ways: for how many of the three lab problems the ``See solution'' feature was used, and how the  overall usage of the tool was distributed over the 9 day period from release of the lab until the deadline.  The actions of interest were: pressing the ``See solution'' button, and making a code submission (either a compilation request or a request to run code against the instructor test suite).  We also calculate the course mark for each student based on the invigilated assessments in the course (i.e., the average of three proctored pen-and-paper exams: two midterms and one cumulative exam), and group students into quartiles for the purpose of our analysis.  

\subsubsection{Qualitative Data}
We asked students an open-ended question after they completed the three lab tasks: 
\textit{If you used the ``See Solution'' feature to generate a code solution for any of the three problems, explain your rationale for using this feature. Reflect on the usefulness of this feature and the extent you used the generated solution in your final submission for the respective problem. Alternatively, if you did not use this ``See Solution'' feature, comment on why you didn’t use it}. 

A total of 839 non-blank responses (Response Rate: 95\%) to this question were analyzed by a single researcher using a reflexive thematic analysis approach~\cite{braun2006}. The process involved 1) the researcher familiarized themselves with the responses, 2) they open-coded the responses, and 3) iteratively identified themes. The themes were selected based on their importance and relevance to the research question. Rather than focusing on coder agreement or inter-rater reliability (which are not appropriate for the reflexive nature of thematic analysis~\cite{mcdonald2019reliability}), our process ensures validity by reflexively engaging with the data by revisiting it multiple times. The researcher actively and iteratively constructs their understanding of the responses, rather than seeks an objective `truth' in the data. Finally, to exemplify and contextualize the themes, we present representative examples from the students' responses. 

\section{Results}

\subsection{Quantitative Analysis}

\subsubsection{Solution-seeking and course performance}
\begin{figure}[h!]
    \centering
    \captionsetup{justification=centering}
    \includegraphics[width=0.95\linewidth, alt={Graph showing usage of ``See solution'' feature by course performance quartile}]{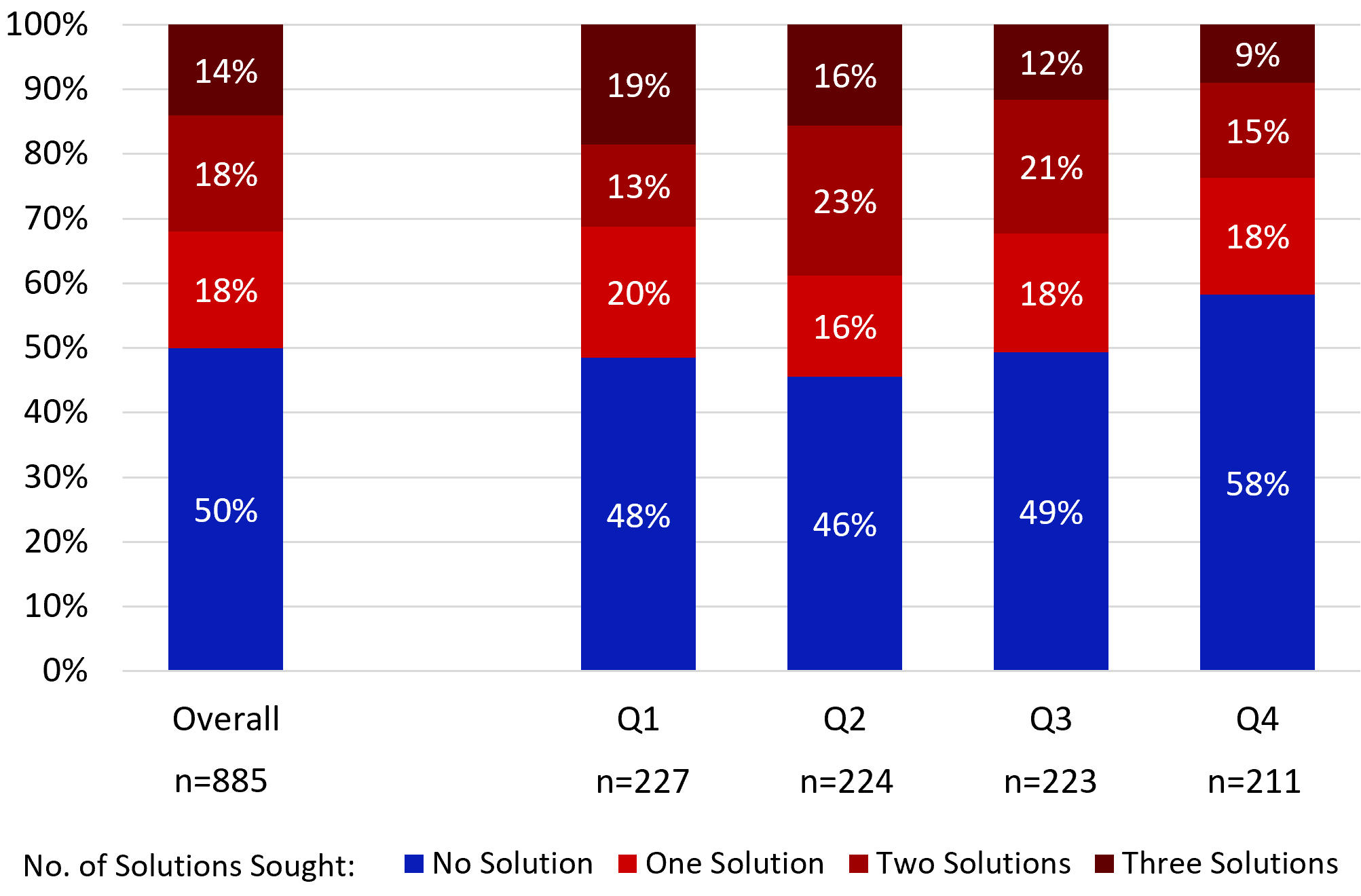}
    \caption{Usage of ``See solution'' feature by course performance quartile (Q1=low, Q4=high)}
    \label{fig:result2}
    \Description {A bar chart titled ``Usage of `See Solution' Feature by Course Performance Quartile'' is presented, with performance quartiles labeled as Q1 (low), Q2, Q3, and Q4 (high) on the x-axis and the percentage or frequency of usage on the y-axis. Each bar represents the proportion of students in the corresponding quartile who used the ``See Solution'' feature, highlighting differences in behavior across performance groups.}
\end{figure}

Students clicked the ``See Solution'' button 1049 times in total across the three problems (ranging from 0 to 10 clicks per student, as some students pressed the button more than once for a problem). 
We observed that 50\% of the 885 students (n=445) did not use the ``See solution'' feature for any of the three problems (see Figure \ref{fig:result2}). Of the remaining 50\% of students who used the feature at least once (n=440), 18\% used the feature for one of the problems (n=160), 18\% used the feature in two of the  problems (n=158), and 14\% used it in all three problems (n=122). 
Across problems, 39\% of students used the feature in the \textit{isPrime} problem (n=349), 17\% used it in the \textit{isRepeated} problem (n=150), and 38\% students used it to see the solution to the  \textit{surroundingSum} problem (n=339).

Students who used the ``See Solution'' feature for all three problems scored, on average, lower on the invigilated tests ($\mu_3=66$) compared to students who did not use the feature ($\mu_0=73$) or used it to see solutions to one or two of the problems ($\mu_1=72$, $\mu_2=73$).



Figure \ref{fig:result2} shows the frequency of use of the ``See solution'' feature across performance quartile ($Q_4$ = highest performing). 
Students in $Q_4$ were more likely to \emph{not} use the ``See Solution'' feature and were less likely to use it in all three problems compared to students in lower quartiles ($Q_1$, $Q_2$, and $Q_3$). For instance, a student in $Q_1$ was twice as likely to use it in all problems than a student in $Q_4$.




\subsubsection{Usage of ``See Solution'' feature over time}

We observed that more students start the lab closer to the deadline as a third of the students used the Submit/Run button to test their code in our system for their first lab task a day before the deadline (see Table \ref{table:solution_by_dates}). However, their decision to use the ``See Solution'' feature was independent of when they started the lab, as approximately 50\% of the students would use the feature regardless of when they started working on the tasks ($\chi^2(7) = 5.92$, $p = 0.55$ indicating no significant differences in start dates of the two groups -- those who used the ``See Solution'' and those who did not). One exception to this were students who started the lab very early (the day the lab was available to them---as 65\% of those students did not use the solution feature).

\begin{table}[ht!]
\centering
\scalebox{0.75}{
\begin{tabular} {|c|c|c|c|c|c|c|c|c|}  
\hline
&  \multicolumn{8}{c|}{Date when a student started the lab (Deadline: 7-Oct)}\\ \cline{2-9}
  Used & 30-Sep & 1-Oct & 2-Oct & 3-Oct & 4-Oct & 5-Oct & 6-Oct & 7-Oct \\
  ``See Solution''? & n=54 & n=88 & n=47 & n=18 & n=124 & n=125 & n=169 & n=258 \\
 \hline
No & 65\% & 50\% & 49\% & 50\% & 46\% & 48\% & 51\% & 50\% \\
 \hline
Yes & 35\% & 50\% & 51\% & 50\% & 54\% & 52\% & 49\% & 50\% \\
 \hline
\end{tabular}
}
\captionsetup{justification=centering}
\caption{``See Solution'' feature usage by date of starting the lab (N=883)}
\label{table:solution_by_dates}
\end{table}

\vspace{-22px}

\begin{figure}[ht!]
    \centering
    \captionsetup{justification=centering}
    \includegraphics[width=0.95\linewidth]{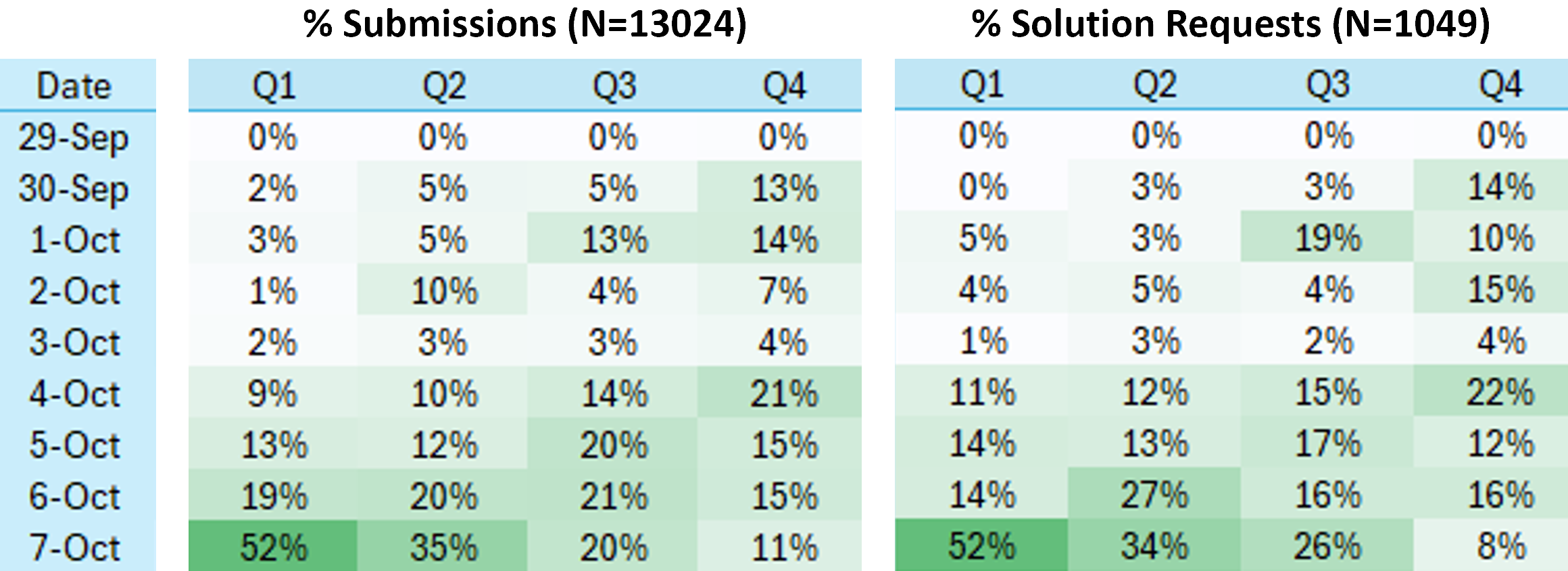}
    \caption{Heatmap of ``See Solution'' feature usage rate and submission rate  \textit{w.r.t.} student performance (N=885 students)}
    \label{fig:result3}
    \Description {Heatmap illustrating the relationship between the usage rate of the ``See Solution'' feature and the submission rate with respect to student performance. The heatmap includes data from $N=885$ students, with student performance categorized into quartiles or performance groups. }
\end{figure}

We also computed the interaction between student performance and ``See Solution'' feature usage. Figure \ref{fig:result3} shows a heatmap of this interaction. We observe that 52\% of the submissions and ``See Solution'' clicks by students in the lowest quartile, $Q_1$ were within the last day before the assignment was due and this behavior close to the deadline gradually decreased from $Q_1$ to $Q_4$. 



These results suggest that low-performing students would procrastinate more compared to high-performing students, but the decision to use the ``See Solution'' feature does not seem to significantly alter with the timing of when students start their labs. Thus, using the ``See Solution'' feature is independent of procrastination, although procrastination is dependent on student performance.

\subsection{Qualitative Analysis}

\subsubsection{Reasons not to view solutions}

Thematic analysis of open response data revealed a variety of reasons why participants chose not to use the ``See Solution'' feature. These reasons included expressing an appreciation for self-directed problem-solving, not needing support, or ethical concerns. 

\textit{Perceived Learning Value and Sense of Accomplishment.} Most of the students who refrained from using the solution feature did so because of the intrinsic value they associated with independently solving the problems and perceived learning benefits. For example, P36 emphasized the importance of preserving learning experience: 

\begin{quote}
    \textit{``I did not use the see solution tool because I find that when you see an already completed solution, it takes away the learning experience that comes from crafting a unique solution yourself.'' - P36}
\end{quote}

Similarly, P1018 shared this concern, \textit{``I did not, as I think that would take away from my learning.''} From a constructivist perspective, the prevalence of students wanting to build their own understanding by working independently is promising. 


\textit{Not needing help.} Another theme identified from the students' responses was that the guardrailed chatbot already provided sufficient support with feedback, debugging support, and helped them identify missing edge cases. 
For example:

\begin{quote}
    \textit{``I did not use the see solution feature as I felt the chat bot offered more than enough assistance for each of the problems.'' - P137}
\end{quote}


Many students described how the chatbot allowed them to quickly develop their understanding and debug any problems that arose. For example,  \textit{``The chatbot was helpful in understanding the code and getting to the root of the issues really quickly.''} For many students, this level of support was already sufficient, with one student describing it as a `cheat code'. 


\textit{Ethical Concerns.} While we explicitly permitted the use of the chatbot and the ``See Solution'' feature, some students described using the solutions as cheating: 

\begin{quote}
\textit{``I chose not to use the `See Solution' feature as it felt a bit like cheating and I wanted to figure it out on my own. Overall, once I learnt how to ask the [chatbot] the right questions, I was able to figure it out much more easily.'' 
- P400}    
\end{quote}


Interestingly, students seemed to feel less concerned about interacting with the chatbot with guardrails. P683 describes how using the chatbot was not `cheating' because it didn't reveal the code, but they equated the `See Solution' feature to cheating: 

\begin{quote}
    \textit{``I think [the chatbot] is generally good, because it don't (sic) show any code so technically you can't use the AI to get the answer quickly. However there is no restriction with see solution feature so it's still easy to cheat'' - P683}
\end{quote}


\subsubsection{Reasons to view solutions}

Participants who used the ``See Solution'' feature described various motivations, including problem-solving assistance, time pressure, curiosity, and lack of self-regulation skills. These themes suggest both cognitive and metacognitive factors that influence students' decision-making.  

\textit{Problem-solving assistance.} The most common reason for using the feature was to seek help when students were stuck or needed to verify their work. Several students described the concept of using the solution to work backward and understand the process: 

\begin{quote} 
    \textit{``This feature was used to reveal what the true answer is. It helped in another way by showing the answer so I can work backwards.'' - P308}
\end{quote}

Others used the feature as a last resort when they were stuck after multiple attempts to solve the problem independently or with the chatbot. These behaviors align with models of self-regulated learning, where students monitor their performance and adjust their strategies by seeking help when needed. Similarly, others used it as a spark of inspiration to get unstuck: 

\begin{quote}
    \textit{``If the chat bot wasn't helpful or I wasn't getting it, I used the solution for inspiration. Solution wasn't always accurate tho.'' - P978}
\end{quote}

Finally, students talked about using the solutions to verify their own process. A few students talked about how this saved them time by confirming they were on the right track. Others talked about how seeing the solution could become fixated in their problem-solving process. P801 described their verification process: 

\begin{quote} 
    \textit{``... after asking the bot 5 to 6 questions, it does give me an opportunity to check and see if my thought process was indeed in the right direction. I would definitely say that using the solution straight away would defeat the purpose of such a learning platform.'' - P801}
\end{quote}


\textit{Time pressure.} Time constraints emerged as a recurring theme in students’ decisions to view the solution. Participants cited tight deadlines and competing priorities as reasons for bypassing exploratory learning and opting for the quickest path to completion:

\begin{quote} %

\textit{``I am submitting this lab quite late so I did not really have time to completely rewrite and debug the code myself so I clicked see solution. [However, without the time pressure] I would genuinely give it a go before wanting to peak at the see solution option.'' - P277}

\end{quote}


Students also used it to save time or to avoid wheel spinning~\cite{beck2013wheel}:

\begin{quote} 
    \textit{``I used the See Solution ... because I wasn't sure how to do 'boundary checks' on indexed elements and I didn't want to spend too much time figuring out a way to do them. For the first two questions, I didn't use the see solution feature as I fully understood what was required...'' - P161}
\end{quote}

\textit{Lack of Self-Regulation Skills.} Some students quite candidly talked about their lack of self-control when it came to using the solution feature. Where some students described being tempted when facing the time pressure of a deadline, others candidly talked about their lack of self-regulation as a primary factor in their decision to use the feature. For example, P941 said, \textit{``I honestly just got lazy, but when it was more straight forward I would not use [the ``See Solution''] feature.''} Similarly, P181 stated `\textit{`don't give us the option to be lazy.'}' These admissions reflect gaps in self-control and intrinsic motivation, which are critical components of self-regulated learning.

\textit{Curiosity.} The final motivation shared by students was curiosity. A few participants used the solution to assess it's accuracy or to compare their correct solution to the solution offered by the tool: 

\begin{quote} 
    \textit{``I used the see solution to see if it generated the correct solution, as I was curious to what it would produce... it was useful to show other ways to solve the code in an efficent manner.'' - P681}
\end{quote}

\section{Discussion}

Contrary to previous work where students expressed their desire to independently arrive at a solution rather than being provided with solutions directly from AI TAs \cite{denny_ai_ta_desireable}, 50\% of students in our study used the ``See Solution'' feature. However, this behavior aligns with student preferences for receiving answers when working with human TAs \cite{lim2023student}.  
Our work provides insight into why students may elect to use such a feature.

\vskip 3pt

\noindent \textbf{Reported Benefits of Solutions.} Based on our qualitative findings, some students reported disabling guardrails to help their learning rather than in order to copy solutions directly. Some students reported exploring solutions to verify their work after completion, while others reported using it to explore alternative solutions.

\vskip 3pt


\noindent \textbf{Reported Concerns over Solutions.} Some students felt they used the solution feature because they lacked self-regulation and self-control.  This raises some concerns about over-reliance on the feature, which aligns with recent work on GenAI use \cite{zastudil2023generative, hou2024effects,prather2023robots,vadaparty2024cs1}.

\vskip 3pt

\noindent \textbf{Potential Impact on Learning.} Students who performed worse in the course were more likely to use the solution feature when solving each task in our study. Although we cannot determine causality from this correlative study in a single lab of a course, it seems likely that struggling students may have been more apt to need to use the solution feature to solve the assignment than higher performing students. Concerningly, if struggling students were to become dependent on using solutions, it could cause an over-reliance on the feature that might hinder learning \cite{hou2024effects, denny_computing_education_cacm, vadaparty2024cs1}. 

\vskip 3pt

\noindent \textbf{Student Procrastination} 
The results of our study also confirm that low-performing students are more likely to procrastinate, similar to the findings of previous work \cite{edwards2009comparing, zhang2022procrastination, liao2019behaviors}. However, the students' decision to use the ``See Solution'' feature appeared to be independent of whether students procrastinated or not.
Qualitative feedback from students indicated that time pressure was a contributing factor to their decision to view the solution, as found in prior work~\cite{lim2023student}.

\vskip 3pt

\noindent \textbf{Future Studies.} Future research could extend this analysis across a course term to draw stronger conclusions. If low-performing students consistently rely on the solution feature over longer periods, their interactions with systems that allow disabling guardrails could serve as an indicator to identify at-risk students who may lack self-regulation skills. Additionally, future research could explore the relationship between effort exerted and solution-seeking behaviors. For example, AI TAs could consider granting access to solutions if students remain stuck on a problem for an extended period of time.

\vskip 3pt

\noindent \textbf{AI TA Design.} Our findings may have implications for designers of AI TA tools. Students reported conflicting opinions about the use of the ``See solution'' button with some reporting they used it to help their learning and others believing it might hinder their learning.  
As such, further evidence is needed to determine whether such a feature can be helpful, specifically under what conditions and for which students it can be helpful. 

\vskip 3pt

\noindent \textbf{Limitations.}  This study was conducted in one lab that occurred late in an introductory programming course at a large research-focused university.  Findings may not generalize outside this context. Additionally, student use of the solution feature may have been influenced by the lab's requirement to interact with a guardrailed AI TA. Students could have been frustrated with the AI TA because it did not receive the problem description as a part of the prompt. This may have encouraged students to rely on the ``See Solution'' feature. Without the AI TA having problem context, students had to write meaningful prompts to receive effective help. Future studies could explore how providing the problem context to the AI TA might affect outcomes of this study and also identify if the AI generated feedback was actually accurate and error free.

Another limitation is students' access to the lab task for over a week, which may have allowed them to consult external resources such as ChatGPT or seek assistance from their peers. While it is hard to determine how external influences affected student decisions, the low stakes of the Edugator task (10\% of lab grade or 0.1\% of course grade) and the option to use the ``See Solution'' button without penalty may have reduced the incentive for misconduct. Future studies in more controlled environments could help isolate the effects of the intervention more clearly. However, our current study allowed us to scale the intervention to a larger cohort of students than would be feasible in a controlled setting.

\section{Conclusion}
In this study, we explore the behavior and motivations of students learning programming when using an AI TA with optional guardrails.
Students reported removing the guardrails from the AI TA for reasons beyond directly copying solutions, such as assistance in problem solving or curiosity to explore alternative solutions. We also found that low-performing students were more likely to request solutions for every problem in the lab assignment. Lastly, we found low-performing students tend to procrastinate and complete the lab closer to the deadline; however, their preference to seek solutions was independent of this procrastination behavior. Our findings contribute to a broader understanding of how students interact with AI tools in education with implications for designing AI TAs that can support educational goals and student motivations.

\bibliographystyle{ACM-Reference-Format}
\bibliography{8_references}

\end{document}